\documentclass[preprint]{vgtc}               

\ifpdf
  \pdfoutput=1\relax                   
  \usepackage{graphicx}                
  \DeclareGraphicsExtensions{.pdf,.png,.jpg,.jpeg} 
\else
  \usepackage{graphicx}                
  \DeclareGraphicsExtensions{.eps}     
\fi%

\graphicspath{{figures/}{pictures/}{images/}{./}} 

\usepackage{microtype}                 
\PassOptionsToPackage{warn}{textcomp}  
\usepackage{textcomp}                  
\usepackage{mathptmx}                  
\usepackage{times}                     
\usepackage{cite}                      
\usepackage{tabu}                      
\usepackage{booktabs}                  

\onlineid{1166}

\vgtccategory{Research}

\vgtcinsertpkg

\title{PRAGMA: Interactively Constructing Functional Brain Parcellations}

\author{Roza G. Bayrak\thanks{e-mail: roza.g.bayrak@vanderbilt.edu}\\ %
        \parbox{1.12in}{\scriptsize \centering Vanderbilt University} %
\and Nhung Hoang \\ %
     \parbox{1.12in}{\scriptsize \centering Vanderbilt University} %
\and Colin B. Hansen \\ %
     \parbox{1.12in}{\scriptsize \centering Vanderbilt University}
\and Catie Chang \\ %
     \parbox{1.12in}{\scriptsize \centering Vanderbilt University}
\and Matthew Berger \\ %
     \parbox{1.12in}{\scriptsize \centering Vanderbilt University}}

\teaser{
 \includegraphics[width=\linewidth]{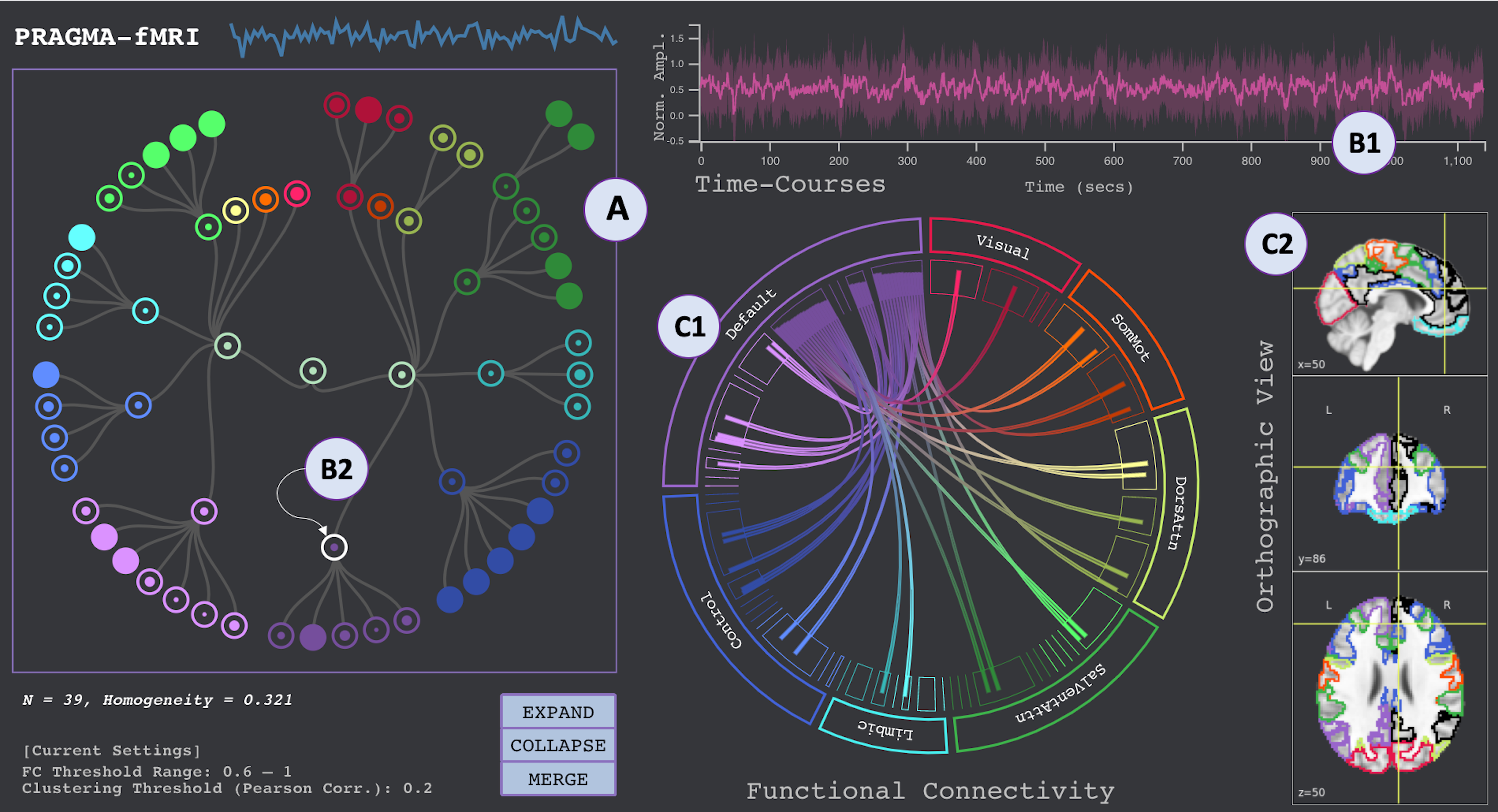}
 \caption{PRAGMA is an interactive tool for constructing scan-specific brain parcellations from mainstream atlases. Its interface incorporates five complementary visuals: (A) abstract view of the hierarchical clustering of the brain through a node-link diagram, (B1) similarity of fMRI signal time-courses represented with a line plot with a shaded confidence interval, (B2) homogeneity glyphs within nodes, (C1) functional connectivity depicted as a chord diagram and (C2) anatomical location of parcels viewed on an orthographic template.}
 \label{fig:pragma}
}

\abstract{A prominent goal of neuroimaging studies is mapping the human brain, in order to identify and delineate functionally-meaningful regions and elucidate their roles in cognitive behaviors. These brain regions are typically represented by atlases that capture general trends over large populations. Despite being indispensable to neuroimaging experts, population-level atlases do not capture individual differences in functional organization. In this work, we present an interactive visualization method, PRAGMA, that allows domain experts to derive scan-specific parcellations from established atlases. PRAGMA features a user-driven, hierarchical clustering scheme for defining temporally correlated parcels in varying granularity. The visualization design supports the user in making decisions on how to perform clustering, namely when to expand, collapse, or merge parcels. This is accomplished through a set of linked and coordinated views for understanding the user's current hierarchy, assessing intra-cluster variation, and relating parcellations to an established atlas. We assess the effectiveness of PRAGMA through a user study with four neuroimaging domain experts, where our results show that PRAGMA shows the potential to enable exploration of individualized and state-specific brain parcellations and to offer interesting insights into functional brain networks.} 

\CCScatlist{
  \CCScatTwelve{Human-centered modeling}{neuroimage analysis}{functional parcellation}{brain mapping}{}
}

\begin{document}


\firstsection{Introduction}

\maketitle

Brain mapping research aspires to capture the complexity of brain organization and the functional roles of brain regions and networks. Atlases are often used by neuroimaging researchers to partition the brain into functionally meaningful units, typically referred to as \emph{parcellations}. Atlases extracted using functional magnetic resonance imaging (fMRI) data are widely used to define nodes for network and functional connectivity (FC) analyses, to simplify a large set of voxels into a small set of regions that are easy to comprehend, and to boost signal-to-noise ratio by averaging across voxels that share temporally homogeneous activity. When derived at the population level, atlas regions are designed to correspond across subjects, thus capturing functional brain patterns that are shared in common across individuals (e.g. we can say that the hippocampus reflects memory-related activity for a large number of individuals). However, there can often be considerable variability in the functional subdivisions of different individuals \cite{fornito2013graph, preti2017dynamic, gonzalez2018task}. Recent research suggests that obtaining neurobiologically meaningful functional regions at the level of the individual might serve as a fingerprint of human cognition or behavior and allow for the investigation of individualized structure-function relationships, psychological traits, or genetic variations \cite{finn2015functional, wang2015parcellating, salehi2018exemplar, bijsterbosch2018relationship, kong2019spatial}.

A range of automated fMRI parcellation techniques have been developed for deriving individual parcellations \cite{wang2015parcellating, gordon2016generation, gordon2017individual, chong2017individual, kong2019spatial}. However, by returning only the end-result of the parcellation, valuable information about relationships between brain regions, and how they are embedded into larger functional networks, is not readily accessible to the user. The ability to interactively parcellate a dataset, while engaging with local and long-range information at multiple spatial granularities, would provide users with richer information about individual differences in functional organization and allow for tailoring the parcellation to the specific research goals. Because the goal behind parcellating fMRI data changes between neuroimaging studies (i.e. clustering voxels based on \emph{some} shared temporal behavior), there is no one formalism for clustering. To this end, we introduce PRAGMA, a visual analytics tool for interactively deriving parcellations of brain regions from fMRI data. PRAGMA depicts fMRI time-courses, and correlations between these signals, to support the user in interactively constructing brain parcellations. PRAGMA is inspired by studies that analyze submodularity of established atlases \cite{wang2015parcellating, gordon2016generation, gordon2017individual}, while following the conceptual framework proposed by Sacha et al.\cite{sacha2017you} for model steering.

However, reliably knowing what patterns to detect in order to make decisions on clustering is a non-trivial problem. To address this challenge, PRAGMA contextualizes the user's clustering with respect to an established atlas, in order to ensure their clustering decisions lead to anatomically and functionally consistent parcellations (see \autoref{fig:pragma}). PRAGMA initializes the parcellation scheme from such an atlas, based on a user-specified Pearson correlation distance, and visually encodes the parcellation as a hierarchical node-link diagram (A). The user can modify their parcellation through this view by expanding, merging, and collapsing nodes, where the decisions that they make are supported by a set of linked and coordinated views that support: time-course data analysis for understanding the variability of time-series data (B1); homogeneity of parcels (B2); a functional connectivity chord diagram to investigate inter-parcel relations (C1); and an orthographic, slice-based view of their current parcellation to locate unique parcels (C2).

To evaluate PRAGMA, we performed a qualitative assessment with four domain experts. Our results demonstrate that PRAGMA could potentially leverage subject-specific analyses by allowing experts interactively identify imprecise regions and reconstruct unique functional brain parcellations.

\section{Related Work}

In this section, we review visualization methods designed for interactive clustering, and visual analysis methods for neuroimaging.

\subsection{Interactive Clustering} There is rich literature on visual exploration of clustering \cite{kwon2017clustervision, l2015xclusim, schreck2009visual, yang2020interactive, das2020geono} and methods that leverage unsupervised clustering \cite{bruneau2015cluster, kern2017interactive, sacha2017somflow, cavallo2018clustrophile}. More recent methods demonstrate the advantage of providing a feedback loop to the domain expert, making clustering human-centered. Hierarchical Clustering Explorer \cite{seo2002interactively} is a pioneering bio-informatics system for the interactive discovery of patterns. Since then, many approaches have been proposed for augmenting interactive clustering, by the means of finding the right clustering algorithm and parameters, and supporting exploratory clustering with visual and statistical analysis \cite{l2015xclusim, kwon2017clustervision, cavallo2018clustrophile, ruta2019sax}.

Akin to PRAGMA, in the aforementioned work, views and computational techniques are combined to help users interactively reach satisfactory clustering results.  In order to use the aforementioned methods, experts need to translate their analytic goals into supported analysis techniques. In contrast, PRAGMA offers the ability to build upon an existing atlas and is equipped with domain-focused cluster analysis techniques that inform the iterative process of parcellating fMRI data into functionally consistent regions. The visualization design supports the user in making decisions on how to perform parcellation, through a set of linked and coordinated views.

\subsection{Visual Analytics in Neuroimaging} Similarly, a wealth of systems have been proposed to visualize and analyse anatomical and functional relationships between brain networks. Network-based neuroscience visualization tools have demonstrated how to gain insight from connectome datasets, some at the level of functional modules, e.g. NeuroCave \cite{keiriz2018neurocave} and TempoCave \cite{xu2019tempocave}, and some at the level of individual neurons, e.g. BrainTrawler \cite{ganglberger2019braintrawler}. A different approach to analyzing functional modules using NodeTrix~\cite{henry2007nodetrix} representations is proposed by Yang et al. \cite{yang2016blockwise}. VisualNeuro\cite{jonsson2020visualneuro} offers a tool for hypothesis formation and reasoning about cohort study data. Brain Modulyzer \cite{murugesan2016brain} is an interactive visual exploration tool for investigating regional correlation and hierarchical network structures.

PRAGMA takes inspiration from the above visual analytics methods, but differs in task. While many methods focus on analysis of pre-defined nodes, PRAGMA focuses on redefining nodes and helping domain experts make parcellation decisions.

\section{Design Objectives and Tasks}

Functional magnetic resonance imaging (fMRI) detects time-varying changes associated with blood flow and captures volumetric time-courses of brain activity. As fMRI data is often noisy and of high spatial and temporal resolution, deriving individual parcellations that reflect meaningful functional information is challenging for domain experts. Automatic clustering methods, e.g. k-means, might yield results that are inconsistent with their prior knowledge, while manually creating parcellations can be a tedious process. Thus, the main objective of PRAGMA is to support domain experts in the analysis, and creation, of scan-specific parcellations.

To derive design objectives for PRAGMA, we initially conducted multiple group discussions with three experts, from the fields of neuroimaging, network and cognitive neuroscience. We started our discussions by asking them about the current practices, needs, and challenges of delineating functional regions for their work. The general direction of the questions were: Do you currently use any established atlases or voxel clustering methods in your analyses? What are the pitfalls and challenges of these processes? How do you evaluate the parcels obtained by clustering? Based on the responses from the experts, we identified a set of high-level design objectives, and refined them as we iteratively received feedback during different stages of development. Our discussions led to the following objectives:

\textit {(DO1) Leverage established atlases in the creation of scan-specific parcellations}: Due to the difficulties of clustering fMRI data, the use of established, meaningful atlases can help domain experts avoid spurious clustering decisions that the clustering algorithm can impose on the data.

\textit {(DO2) Provide an intuitive approach for modifying the granularity of parcels}: Domain experts should be able to decide the granularity of parcels based on the need of the study, and apply their domain knowledge for merging similar regions together and splitting inhomogeneous regions further into smaller regions. 

\textit {(DO3) Provide useful within- and between- parcel information that supports parcellation decisions}: The design should provide access to information about the current parcel properties, so that modifications are justifiable.

From these objectives, we list the following tasks that our visualization design aims to support:

\textit{(T1) Steer parcellations using a population-based atlas (DO1)} The design should support the use of established atlases, both for initializing the clustering, and as a guide for steering clustering.

\textit{(T2) Support inter-parcel comparisons (DO3)} The design should support the user in understanding the relationship between two parcels in their current parcellation.

\textit{(T3) Support intra-parcel comparisons (DO3)} Users should be able to assess the homogeneity of a given parcel, e.g. how similar are a set of time-courses that belong to a single parcel. 

\textit{(T4) Steering parcellation (DO2)} Our design should allow the user to interactively modify the parcellation, where we identify three basic operations for steering: (1) splitting a parcel, (2) collapsing a set of parcels into one, and (3) merging two parcels.

We emphasize that tasks \textit{T1-3} are intended to provide guidance for the user in making decisions on steering \textit{(T4)}. Namely, a parcel should be split if the user does not consider it homogeneous, two parcels should be merged if they are sufficiently similar, and likewise, a set of parcels should be collapsed into one if they are all considered similar.

\section{Visualization Design}

In this section, we describe the visual encodings of the PRAGMA interface and their interactions. The interface is composed of a Hierarchical Node-Link Diagram (A), two Parcel Specific Views (B),  and two Current Parcellation Views (C) (see \autoref{fig:pragma}).

\subsection{Hierarchical Node-Link Diagram (A)}\label{HTD}
The diagram represents the full hierarchy of the parcellation, and it is the main functional visual in the tool (see \autoref{fig:pragma}). The parcels are grouped by functional networks in the left and right hemispheres. Each network is represented by a unique color, and this coloring is consistent across the node-link diagram, chord diagram, and orthographic views. Furthermore, lighter and darker hues represent networks in the left and right hemispheres respectively. While the root node represents the whole brain, each leaf node in the hierarchy represents a parcel from the current parcellation. 

\begin{figure}[tb]
 \centering 
 \includegraphics[width=\columnwidth]{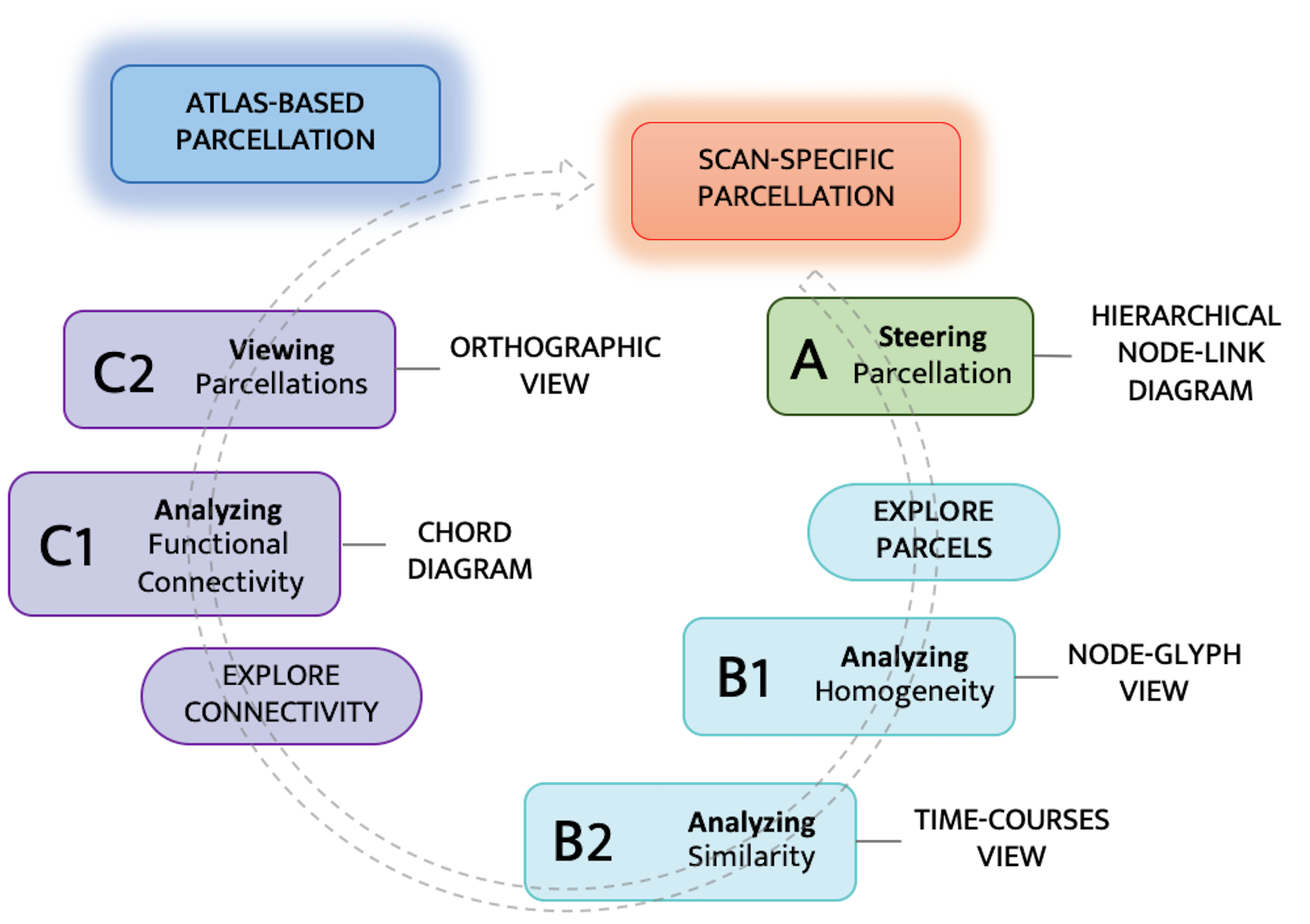}
 \caption{Iterative process of creating scan-specific parcellations from mainstream atlases. Each view provides some useful information to support and facilitate the decision on parcellation.}
 \label{fig:iterative}
\end{figure}

The user selects a distance threshold to initialize the hierarchical node-link diagram. Pearson's correlation is used to form a precomputed distance matrix based on this threshold. The initial node-link diagram is formed by applying complete-linkage agglomerative clustering to this precomputed matrix, done on a network-by-network basis. Afterwards, the user may explore parcels by selecting nodes, which updates the remaining views with parcel-specific information (see \autoref{fig:iterative}). Selecting a node will return the aggregated time-course data encoded as mean +/- standard error, functional connectivity of the selected parcel(s) to the other parcels, and anatomical locations of these regions in the orthographic view. 

\begin{figure}[tb]
 \centering 
 \includegraphics[width=\columnwidth]{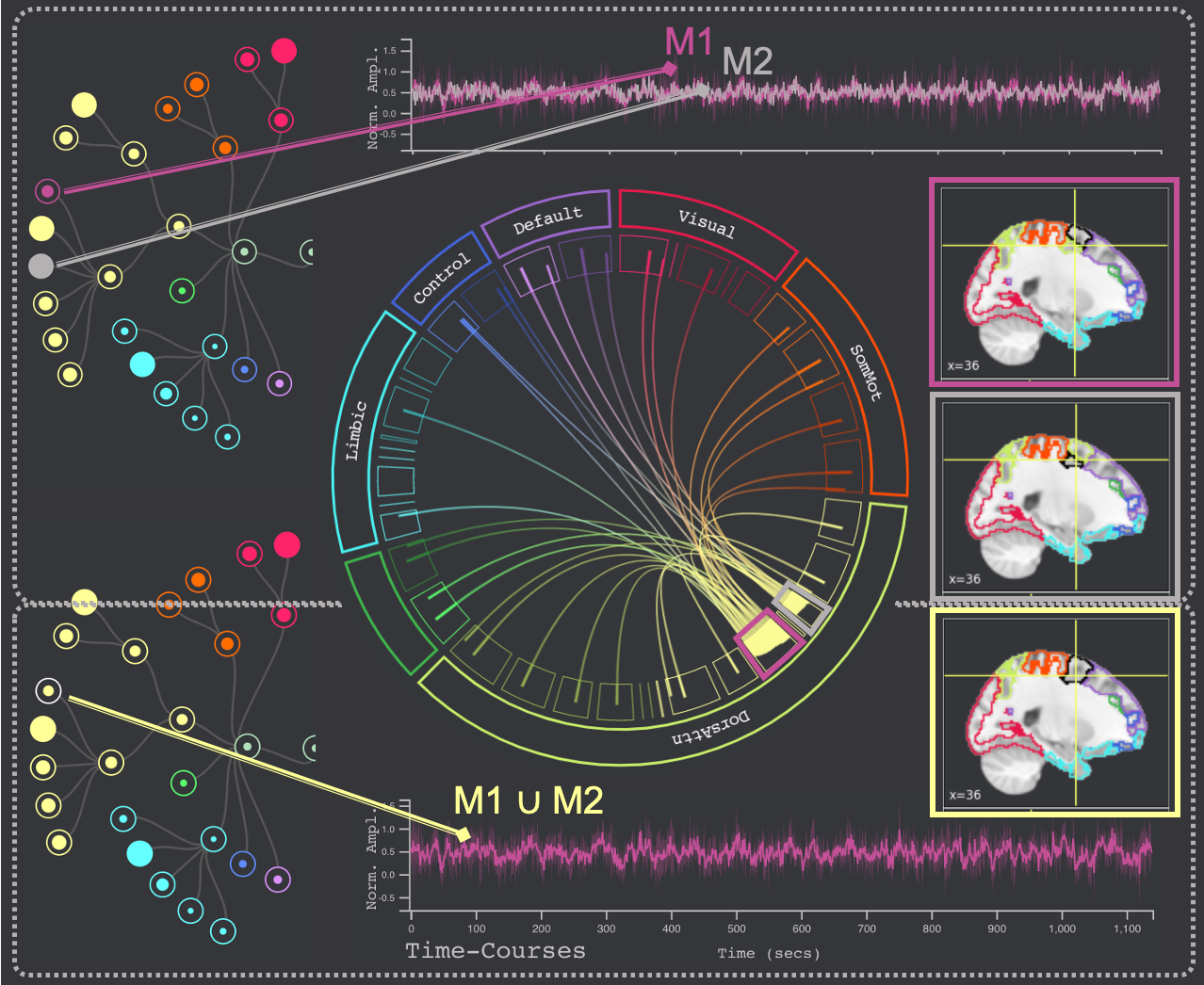}
 \caption{To merge two parcels together, a user selects node (M1) followed by node (M2). The top time-courses plot shows the two signals overlaid on top of each other for comparison. After the \emph{merge} is applied, the super-voxels from the second node (M2) are merged with the first node (M1). The time-courses plot is updated to reflect the merge, along with the orthographic view.}
 \label{fig:merge}
\end{figure}

In case two or more clusters are found to be similar, we support a \emph{merge} action \textit{(T4)} (see \autoref{fig:merge}). To investigate such regions, a user can double click on a leaf node of interest. This will lock the parcel-specific information, and it will persist in view unless the node is released. Users can then continue exploring other parcels in reference to this selected parcel by single click selecting other nodes. Once the user finds another parcel that they want to merge into the initially selected node, they select it with a double click on this second leaf node. The user may then click on the \textit{MERGE} button to perform the merging of these nodes, and parcel-specific views are subsequently updated to show the newly merged region. 

While \emph{merge} only allows two regions to form into one, \emph{collapse} can combine all the leaf nodes into a non-leaf node. A user can first select a non-leaf node and then click on the \textit{COLLAPSE} button to demonstrate that this is the intended action. The parcels can not be collapsed into the root, hemisphere, or network nodes, in order to keep the hierarchy intact. However, parcels can be identified and merged into different networks. 

If users deem that within-parcel similarity is low, they can further split this parcel into smaller regions. Subsequent clustering requires a more constricting distance threshold. After this threshold is selected, users click on the \textit{EXPAND} button to demonstrate that this is the intended action.

\subsection{Parcel-Specific Views (B)}
This portion of the visualization is designed to provide real-time analysis of within-parcel properties, as a means of evaluating parcels and supporting parcellation decisions.

\subsubsection{Intra-Parcel Time-Series Similarity (B1).}
We designed a mean time-course line plot with a standard error confidence interval. This visual encoding demonstrates the similarity of the aggregated time-courses of all super-voxels grouped in a selected parcel. The main use of this analysis is to evaluate if a parcel should be further expanded. Additionally, the user can overlay the time-courses of two parcels for comparison. Brush selection is also supported to allow the user to zoom in on the time axis.

\subsubsection{Parcellation Homogeneity (B2).}
Homogeneity is defined as average pairwise temporal correlation, calculated using Pearson's correlation. This quantitative information is useful for making decisions about whether a parcel needs to be further divided into finer parcels. We carry this information within the node-link diagram, encoding homogeneity within the inner circle of each node. The homogeneity is normalized to the radius of the circle, where a more homogeneous parcel is represented by a bigger circle. Homogeneity is calculated for the entire hierarchy, allowing the user to compare each parcel in relation to the parcels it is linked to.

\subsection{Current-Parcellation Views (C)}
This portion of the visualization is designed to support analysis of inter-parcel properties for the current set of non-leaf nodes.

\subsubsection{Functional Connectivity (C1).}
The chord diagram communicates functional connectivity (FC) patterns between the current set of leaf nodes. Functional connectivity describes co-activity of discrete regions in the brain and is calculated as the correlation between every pair of regions in the current parcellation. The inner arcs represent the parcels, and the chords encode the presence of connectivity between parcels. Each inner arc is accompanied by a bar plot encoding the strength of connectivity between its corresponding parcel and the other parcels. The connectivity strength is normalized to the height of each arc, where a stronger FC is represented by a higher bar. The width of the arc represents how connected a parcel is to other parcels in the current scheme. In the chord diagram, the parcels are nested under the functional networks, which are labelled in the outer arcs. When a node is selected from within the node-link diagram or an inner arc is hovered over, the chord diagram illuminates the connections of just the corresponding parcel. To reduce visual clutter arising from the large number of connections, we allow the user to filter out chords within a prescribed range of FC strength.

\subsubsection{Orthographic Parcellation View (C2).}
The orthographic view shows the current parcellation scheme mapped onto a template image and highlights the corresponding physical locations of these parcels (see \autoref{fig:pragma}). Every parcel is outlined by the color representing its functional network. The provided sliders allow users to view the current functional parcellation at different sagittal (yz), coronal (xz), and axial (xy) planes. When a parcel is selected in the node-link diagram, the corresponding parcel contour in the orthographic view is outlined in black.  

\section{User Study}
A user study was performed to investigate the effectiveness of the tool. Four domain experts (two faculty, two post-docs) who have expertise in the understanding of functional networks evaluated the tool. Prior to the study, at different stages of development, three of these domain experts provided feedback that was used in the iterative process to improve the tool and one is a co-author in the paper. The participants received no compensation for taking part in the study. 

\textbf{Data.} We used a single subject resting-state (rs-fMRI) scan from the Human Connectome Project \cite{van2013wu} and Schaefer cortical brain atlas derived from rs-fMRI data \cite{schaefer2018local}. The rs-fMRI scan has been pre-processed with the “FIX” de-noising pipeline\cite{griffanti2014ica} and has undergone the HCP “minimal preprocessing pipeline” \cite{glasser2013minimal}. A Schaefer atlas is applied to this subject scan to initialize super-voxels by extracting averaged time courses within parcels. The latter process reduces the spatial dimension from 200K voxels to 400 super-voxels.

\textbf{Implementation Details.} The front-end of our tool was developed as an Observable notebook, enabling ease of access for our study, alongside a Python server that supports real-time analyses and computation. We have made PRAGMA publicly available at: https://github.com/neurdylab/PRAGMA.

\textbf{Zoom-proceedings.} The domain experts were invited to individual Zoom-meetings. They had been provided a user guide that included (1) instructions to install PRAGMA, (2) pre-/post- study surveys, and (3) written/animated instructions on how to use the tool. First, they were asked to fill the pre-study survey. Then the experts were explained the basic actions that they can take to use the interface (brush, hover over, etc.), as well as the visual encodings and their interactions. They were provided the pre-processed scan and asked to explore the data. The exploration of the tool was recorded via screen-record over Zoom. During this meeting they were informed that they could ask design/functionality related questions to the visual analytics researchers. There were no restrictions on how they could use the tool, nor any time restrictions. For the group of experts who were more familiar with fMRI data parcellation, this freeform exploration took about an hour; for the other group, less than half an hour. Lastly, they were asked to fill a post-study survey to finalize their evaluation.

\textbf{Findings.}
The findings are categorized into three main sections: design choices, intuitiveness, and scope. \textit{Design choices}: Four experts rated the design choices (linked highlighting, spatial arrangements, etc.) as 7.5/10 on average. One expert commented that the arrangement of the parcels in the chord diagram (grouped by network) being different than the node-link diagram" (grouped by hemisphere), increased the cognitive load. On another note, the users observed that being able to compare two parcels on multiple views at the same time and interact with them (i.e. brush to zoom) was helpful. \textit{Intuitiveness}: The average rating of all four experts for ease of use is 6.5/10. Two of the experts commented that they initially struggled with how to engage efficiently with the tool but they quickly picked up even though "there are a lot of features" to support their analyses. While the chord diagram and 2D orthographic views were found to be the most useful, the users suggested that rather than using a slider, they would prefer to be able to click on the image to navigate it.  \textit{Scope}: Two post-docs rated 8/10 that they would use PRAGMA to support subject-specific analyses. On the other hand, one expert suggested that a search for valid cognitive relations through single-scan analysis was likely to result in inconsistencies, mainly because fMRI data is very noisy. One of the experts suggested that a tool like PRAGMA could also be used for population-level exploration and clustering. Two of the experts who explored the tool used a similar sequence of actions: they located a region that they are familiar with in the orthographic view, found this region on the node-link diagram, and then compared the time-series signal and functional connectivity of the corresponding parcel to nearby parcels. The set of actions taken by these users suggested that the prior atlas was helpful.

\section{ Discussion and Conclusion}
Forming precise and meaningful definitions of brain regions is an open and pressing research question. Our approach, PRAGMA, addresses these issues by supporting users in the customization of existing atlases to create individualized parcellations based on their domain needs. Our parcel-by-parcel analysis allows unique parcellations, while ensuring the results remain coherent with established atlases. The user study has shown that PRAGMA offers an intuitive way to understand individual-level depictions of brain activity that are not possible to resolve in group-level atlases. For future work, we plan on addressing issues related to common sources of noise in fMRI data~\cite{liu2017global,zhang2020understanding} through visually conveying uncertainty as part of our design. Furthermore, we plan on making our system more extensible, handling various atlases, clustering algorithms, and similarity measures, in order to better address demands from domain experts.

\acknowledgments{
This work was supported in part by
a grant from NSF IIS-2007444.}

\bibliographystyle{abbrv-doi}

\bibliography{camera-ready}
\end{document}